\documentclass{aastex}
\usepackage{emulateapj5,natbib,makeidx,epsfig,amsmath,epic,tabularx,subfigure}

\begin{document}

\title{A PERSPECTIVE ON THE CMB ACOUSTIC PEAK}
\author{T.A. Marriage\altaffilmark{1,2}}
\altaffiltext{1}{University of Cambridge, Department of Applied Mathematics
and Theoretical Physics, Centre for Mathematical Sciences, Wilberforce Road, 
Cambridge CB3 0WA}
\altaffiltext{2}{Current address: Princeton University, Department of Physics, 
Jadwin Hall, Princeton, NJ 08544}

\begin{abstract}
CMB angular spectrum measurements suggest a flat universe. 
This paper clarifies the relation between geometry and the spherical
harmonic index of the first acoustic peak ($\ell_{peak}$).
Numerical and analytic calculations show that $\ell_{peak}$
is approximately a function of $\Omega_K/\Omega_M$ where $\Omega_K$ and
$\Omega_M$ are the curvature ($\Omega_K > 0$ implies an open geometry) 
and mass density today in units of critical density. 
Assuming $\Omega_K/\Omega_M \ll 1$, one obtains 
$\ell_{peak} \approx \frac{11\sqrt{3}}{9(\sqrt{a_*+a_{eq}}-\sqrt{a_{eq}})}
\left(2+\frac{\Omega_K}{\Omega_M}\right)$ where $a_*$ and $a_{eq}$ are 
the scale factor at decoupling and radiation-matter equality.
The derivation of $\ell_{peak}$ gives another perspective on the 
widely-recognized $\Omega_M$-$\Omega_\Lambda$ degeneracy in flat models. 
This formula for near-flat cosmogonies together with current angular 
spectrum data yields familiar parameter constraints.
\end{abstract}
\keywords{cosmic microwave background --- cosmological parameters}

\section{INTRODUCTION}

It has long been recognized that the first peak in the CMB angular
spectrum provides information about the curvature of the universe
\citep{doroshkevich, bond, kamionkowski_1, efstathiou, cornish, weinberg_1}.
The data are now in. TOCO, BOOMERanG, MAXIMA, and DASI 
have measured the peak position
\citep{amber,netterfield,lee,halverson}.
Analyses of the data strongly suggest a flat universe 
\citep{hu_3,jaffe_1,stompor,pryke,dodelson}. 
The MAP satellite, cosmic variance limited through $\ell \approx 600$, 
should make the definitive measurement in the near future 
\citep{page_1}. 

As new data resolve higher order peaks, attention shifts to the physics
at angular scales beyond that of the first maximum. The object of
the present analysis is to clarify the physics derived
from the position of the first peak, hereafter called ``the peak index'' or
$\ell_{peak}$.
The size of the sound horizon $r_{s*}$ at an angular 
diameter distance $D_{a*}$ to decoupling
determines the peak index.\footnote{An expression subscripted by `*' 
or `eq' is evaluated at decoupling or matter-radiation equality 
respectively.} 
This is widely recognized and serves as a starting point.
In Sections \ref{sec:da}, \ref{sec:snd}, and \ref{sec:angleandindex}, 
numerical and analytic calculations yield the peak index as a 
function of $\Omega_M$ and $\Omega_K$. All results are checked 
with CMBFAST \citep{seljak_CMBFAST}.
In Section \ref{sec:dscsn}, a simple formula for $\ell_{peak}$, applicable to
low $\Omega_K/\Omega_M$ universes, is developed alongside geometric and
classical interpretations of the formal results. It is found that the 
peak index approximates a function of $\Omega_K/\Omega_M$. 
Although our analysis grounds itself in the familiar 
concepts of $D_{a*}$ and $r_{s*}$, the 
results are not widely recognized. While the physical effects
responsible for $\ell_{peak}$ are understood, the interplay that
gives the $\Omega_K/\Omega_M$ dependence is not intuitive and holds 
a number of surprises. We end the investigation by using the functional form
of $\ell_{peak}$ and current angular spectrum data to 
obtain parameter constraints resembling those of more 
sophisticated treatments.

Throughout this work, the ($\Omega_M$,$\Omega_K$) plane serves
as the parameter space. Other quantities enter, though the relation
between $\ell_{peak}$ and $\Omega_K/\Omega_M$ is relatively independent 
of them. To be concrete, we take $\Omega_Bh^2=0.02$, 
consistent with nucleosynthesis \citep{burles}.
The redshift of decoupling, $z_*$, is taken as 1400 from the Saha equation.
The Hubble constant assumes 72 km/s/Mpc in accord with
the HST Key Project results \citep{freedman_1}.

\begin{figure*}
\begin{center}
\centering
\epsscale{1.8}
\plotone{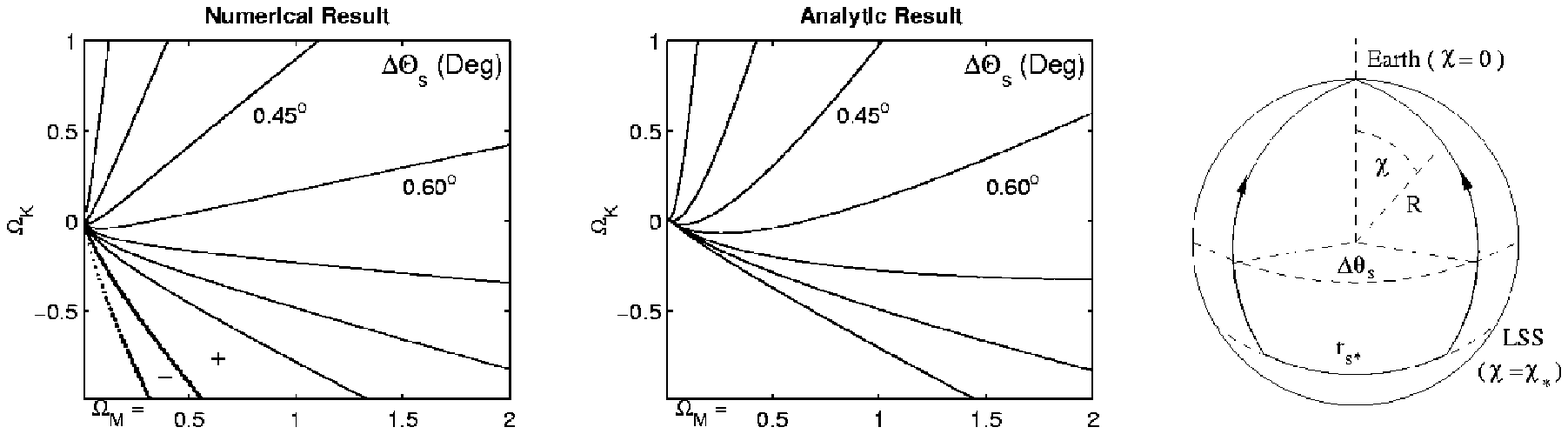}
\caption{\footnotesize
The angle subtended by the sound horizon at decoupling.
$\Delta\Theta_S = r_{s*}/D_{a*}$. Data favors $\Delta\Theta_s = 0.60^\circ$
\citep{knox_2}.
In both numerical and analytic plots, equation (\ref{eqn:hrzn_slvd})
gives the sound horizon $r_{s*}$. In the numerical plot,
the angular diameter distance to last scatter $D_{a*}$
is computed using the cosmography code \citep{hogg_1}.
In the analytic result, the classical equations for $D_{a*}$
are integrated with $\Omega_\Lambda,a_{eq},a_*=0$, and the resulting formulae
are used without assuming $\Omega_M + \Omega_K = 1$
(see eq. [\ref{eqn:daa}]).  $\Delta\Theta_S$ from either method
approximates a function of $\Omega_K/\Omega_M$.
The bending of the contours in the analytic result
arises from radiation-dominated dynamics before last scatter.
In the numerical result, explicit inclusion of $\Omega_\Lambda$
in the Friedmann equation balances the radiation effect:
$\Delta\Theta_S$ contours are straight
lines converging on zero in the full calculation. At far right is
illustrated the comoving coordinates of a closed FRW universe with
time and azimuthal angular coordinates suppressed.
Inspection of the drawing shows
$\Delta\Theta_s=r_{s*}/R\sin\chi_*$ where the comoving sound
horizon at decoupling $r_{s*}$ and
the comoving distance to decompling $R\chi_*$ are calculated
using Friedmann dynamics.}
\label{fig:dths}
\end{center}
\end{figure*}

\section{ANGULAR DIAMETER DISTANCE}
\label{sec:da}

The comoving Friedmann-Robertson-Walker (FRW) metric for
a closed universe ($\Omega_K < 0$) may be written as 
\begin{equation}
ds^2 =  d\eta^2 - R^2 \left( d\chi^2 + \sin^2 \chi d\Sigma^2 \right) 
\label{eqn:frw}
\end{equation} 
where $\eta$ denotes conformal time, $R^2=-1/\Omega_K$, and 
$d\Sigma^2$ is the line element of a unit two-sphere. 
All spacetime intervals are given in units of $H_0^{-1}$.
The angular diameter distance 
$D_a$ to an object of proper length $ds$ at 
comoving distance $R\chi$ ($d\eta = d\chi = 0$) is defined 
so that $d\Sigma = \mid ds \mid /D_a$. 
\begin{equation}
D_a =  R \sin \chi(a)
\label{eqn:rawda}
\end{equation}
where $\chi^2$ is the solution to the Friedmann equation:
\begin{equation}
d \chi ^2 =  \frac{-da^2}{\Omega_M(a/\Omega_K + a_{eq} /\Omega_K)
+ \Omega_\Lambda a^4/\Omega_K + a^2}.
\label{eqn:friedmann}
\end{equation}
Flat and open geometries are treated analogously.

{\bf Numerical Result} $D_{a*}$  can be evaluated numerically. The distance
to redshift 1400 is computed for models across the
$(\Omega_M,\Omega_K)$ plane using cosmography routines from 
David Hogg \citep{hogg_1}. 

{\bf Analytic Result} To complement the numerical result, 
$D_{a*}$ can be estimated analytically by setting 
$\Omega_\Lambda, a_{eq} , a_*$ to zero in equation (\ref{eqn:friedmann}), 
and then using Mattig's solution (cf. \citep{peebles}).
\begin{equation}
D_{a*} = \frac{2}{\sqrt{\Omega_M}}
\times S(\gamma)
\label{eqn:daa}
\end{equation}
where $\gamma=2 \mid\Omega_K\mid/\Omega_M$, and S=1 
for a flat universe. If $\Omega_K$ is non-zero, then S takes the form
\begin{equation}
\begin{array}{c@{\quad}c@{\quad}l}
  S(\gamma) & = & \sqrt{1/2\gamma} \times \left\{
\begin{array}{l@{\quad:\quad}r}
\frac{1}{2}(\beta  - 1/\beta) & \Omega_K > 0 \\
\sqrt{1-(\gamma-1)^2} & - \Omega_M < \Omega_K < 0  \\
\end{array}
\right. \\
 && \\
  \beta & = & \gamma+1+\sqrt{(\gamma+1)^2-1}.
\end{array}
\label{eqn:R}
\end{equation}

\section{THE SOUND HORIZON}
\label{sec:snd}

Sound travels through a tightly coupled photon-baryon 
system with speed $c_s$ 
\begin{equation}
c_s = \frac{1}{\sqrt{3(1+Q)}}
\label{eqn:snd_spd}
\end{equation}
where Q is 3$\rho_B$/4$\rho_\gamma$, and $\rho_B$ and
$\rho_\gamma$ are the energy densities of baryons and radiation,
respectively. 
The sound horizon at decoupling $r_{s*}$ in comoving coordinates is then 
\begin{eqnarray}
r_{s*} & = & \int_{a=0}^{a_*} c_s \frac{d\eta}{da}da \nonumber \\
   & = & \frac{2}{\sqrt{3\Omega_M}}\sqrt{\frac{a_{eq}}{Q_{eq}}}
\ln\frac{\sqrt{1+Q_*} + \sqrt{Q_*+Q_{eq}}}{1+\sqrt{Q_{eq}}}
\label{eqn:hrzn_slvd}
\end{eqnarray}
\citep{hu_2}. Note that curvature does not affect sound dynamics 
before decoupling. In any geometry, $r_{s*}$ will be the same.

\section{THE HORIZON ANGLE AND THE PEAK INDEX}
\label{sec:angleandindex}

$D_{a*}$ and $r_{s*}$ give the angular size of the horizon:
\begin{equation}
\Delta\Theta_{s}=\frac{r_{s*}}{D_{a*}}.
\label{eqn:dths}
\end{equation}
Values for $\Delta\Theta_s$ corresponding to the numerical and 
analytic results for $D_{a*}$ (see Section \ref{sec:da}) are plotted 
in Figure \ref{fig:dths}. Curves of constant $\Delta\Theta_{s}$ 
approximate straight lines, which intersect
the origin of the ($\Omega_M,\Omega_K$) plane. 
\emph{The angle subtended by the sound horizon, and so the position
of the first peak, approximates a function of 
$\Omega_K/\Omega_M$.} 
A well known corollary to this general statement is that a peak
corresponding to $\Omega_K=0$ should be insensitive to 
variations in $\Omega_M$ \citep{bond,hu_3}.

\begin{figure*}
\begin{center}
\centering
\epsscale{1.8}
\plotone{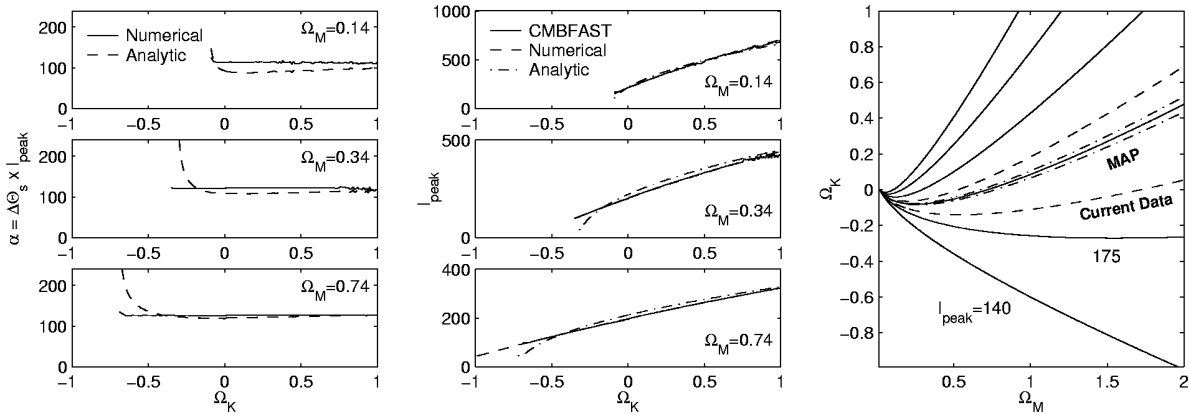}
\caption{\footnotesize
The relation $\alpha\;=\;\Delta\Theta_S \times \ell_{peak}$
and parameter constraints.
The peak index from CMBFAST times $\Delta\Theta_S$ of numerical and analytic
calculations gives $\alpha$ (left frame). The curves for $\alpha$ are
averaged and reinserted into the equation $\ell_{peak}=\Delta\Theta_S/\alpha$
to generate the corresponding graphs of peak index (middle frame).
Contours of $\ell_{peak}$ in the ($\Omega_M$,$\Omega_K$) plane
are calculated using equations (\ref{eqn:rsapprox}) and
(\ref{eqn:daapprox}) (right frame).
To coincide with the values of $\Delta\Theta_s$ and $\ell_{peak}$ derived from
current data, the contour plot assumes
$\alpha = 125 (\approx 0.6^\circ \times 210)$ throughout the plane
\citep{knox_2, knox_1, hu_3}. The dashed contours show constraints from
current estimates of $\ell_{peak}$ and from the projected MAP satellite
measurement. }
\label{fig:cmbfast}
\end{center}
\end{figure*}

To check whether this simple analysis agrees with the standard model, we
compare the above results to those from CMBFAST \citep{seljak_CMBFAST}.
CMBFAST calculates the CMB angular 
spectrum.\footnote{CMBFAST inputs are
$(\Omega_B=0.04,\Omega_\nu =0,H_0=72,T_{cmb}=2.726,Y_{He}=0.24,
N_{\nu}(massless)=3.04,N_{\nu}(massive)=0$,recfast,no reion,
scalar only,primordial index 1,adiabatic).}
Then, $\ell_{peak}$ from the spectrum multiplies
$\Delta\Theta_s$ from Figure \ref{fig:dths} to give the constant of 
proportionality $\alpha=\ell_{peak}\times\Delta\Theta_s$.  As shown in
Figure  \ref{fig:cmbfast},  the numerically derived 
$\alpha$ increases from 110 ($\Omega_M=0.14$) 
to 125 ($\Omega_M=0.74$) and has weak if any dependence on curvature.
The near constant graphs of $\alpha$ suggests a well-defined agreement
between the peak indices from CMBFAST and those derived from this work's
numerical and analytic calculations (see middle frame of 
Figure \ref{fig:cmbfast}).

\section{DISCUSSION}
\label{sec:dscsn}

To gain intuition about $\Delta\Theta_s$, consider the closed comoving
universe (eq. [\ref{eqn:frw}]) illustrated in Figure \ref{fig:dths}, 
where time and azimuthal angular coordinates have been supressed 
to produce the familiar two-sphere geometry. In this picture,
CMB photons follow great circles, and, assuming a small angle,
$\Delta\Theta_s$ derives from inspection:
\begin{math}
\Delta\Theta_s = r_{s*}/R \sin \chi_*,
\end{math}
which is exactly equation (\ref{eqn:dths}).

To better understand the parameter dependence of $r_{s*}$,
take the sound speed $c_s$ (eq. [\ref{eqn:snd_spd}]) to be constant at 
$9/10\sqrt{3}$. (With $\Omega_Bh^2=0.02$ and radiation density
derived  from the COBE FIRAS measurement, the sound speed decreases at
a near constant rate ($dc_s/da \approx constant$) from $c/\sqrt{3}$ at a=0 
to four-fifths that value at decoupling \citep{fixen}.) 
The sound horizon is then 
\begin{equation}
r_{s*} = c_s \int_{0}^{a_*}\frac{d\eta}{da}da = 
\frac{9}{5\sqrt{3\Omega_M}} (\sqrt{a_*+a_{eq}} - \sqrt{a_{eq}})
\label{eqn:rsapprox}
\end{equation}
where the final equality follows from the Friedmann equation 
(\ref{eqn:friedmann}) with $\Omega_K=\Omega_\Lambda=0$. 

To obtain a simple formula for $\ell_{peak}$, expand $D_{a*}$ 
to first order in $\gamma = 2\mid\Omega_K\mid/\Omega_M$:
\begin{equation}
D_{a*} = \frac{1}{\sqrt{\Omega_M}} \left( 2 +
\frac{\Omega_K}{\Omega_M} \right).
\label{eqn:daapprox}
\end{equation}
The expansion applies to models with
a low curvature-to-matter ratio. Coincidentally, such models are
favored by experiment, so that the resulting formula for $\ell_{peak}$
is useful. Given equations (\ref{eqn:rsapprox}) and (\ref{eqn:daapprox}) and
a value of $\alpha$ from Figure \ref{fig:cmbfast},
\begin{equation}
\ell_{peak} = \frac{\alpha D_{a*}}{r_{s*}} \approx 
\frac{11\sqrt{3}}{9(\sqrt{a_*+a_{eq}}-\sqrt{a_{eq}})}
\left(2+\frac{\Omega_K}{\Omega_M}\right).
\label{eqn:roughell}
\end{equation}
This near-flat approximation is plotted in Figure \ref{fig:cmbfast}
where $a_{eq} \approx 2.4 \times 10^{-5} (\Omega_M h^2)^{-1}$. 
For flat models, the distance to last scatter 
(eq. [\ref{eqn:daapprox}]) scales as $\Omega_M^{-1/2}$. 
One may intuit that self-gravitation leads to
smaller cosmological separations. This same ``gravitational'' 
effect decreases the distance between the big bang and last
scatter, and therefore $r_{s*}$ (eq. [\ref{eqn:rsapprox}]) also
scales with an overall factor of $\Omega_M^{-1/2}$.
In equation (\ref{eqn:roughell}), the $\Omega_M^{-1/2}$ dependence of
$D_{a*}$ cancels that of $r_{s*}$. This cancellation
helps explain the $\Omega_M$-$\Omega_\Lambda$ degeneracy of 
flat cosmogonies. Furthermore, the $\Omega_K/\Omega_M$ term in equation
(\ref{eqn:roughell}) is proportional to the curvature times the 
area between light rays in a two-dimensional representation of
an $\Omega_K \approx 0$ universe (e.g. Figure \ref{fig:dths}). 
This suggests that one may think of ($\Omega_M$,$\Omega_K$) dependence of 
$\ell_{peak}$ in near-flat spacetimes as resulting from light curving
like the geodesics of a two-dimensional space of constant curvature. Finally,
the effect of radiation on early universe dynamics and, in particular,
on $r_{s*}$ is made explicit by the appearance of $a_{eq}$ in 
equation (\ref{eqn:rsapprox}). Radiation-dominated
cosmological growth per expansion scale ($d\eta/da$) is less than 
that of matter-dominated dynamics. In low $\Omega_M$ universes,
radiation brings last scattering even closer to the big bang and so
shortens $r_{s*}$. Thus, the effect of radiation in the early 
universe is to spoil the pure $\Omega_K/\Omega_M$ dependence of 
$\ell_{peak}$ as manifest by the curved contours in the plot of 
equation (\ref{eqn:roughell}) in Figure \ref{fig:cmbfast}.
The straightness of the $\ell_{peak}$ contours is restored in the 
numerical result shown in Figure \ref{fig:dths}. 
This suggests that explicit inclusion of $\Omega_\Lambda$ in
the computation of $D_{a*}$ balances the $a_{eq}$ dependence of 
$r_{s*}$.

\section{CONCLUSION}

The  peak index has long been recognized as an 
indicator of geometry. It is hoped that the present 
analysis sheds new light on $\ell_{peak}$. 
The peak index does not determine the magnitude of curvature, 
but rather the ratio of curvature to matter. A measurement of the peak's 
angular scale gives the precise geometry only if $\Omega_K \approx 0$, 
otherwise $\ell_{peak}$ is a function of $\Omega_K/\Omega_M$. 
Furthermore, in deriving the
$\Omega_K/\Omega_M$ dependence of $\ell_{peak}$, unexpected
cosmological cancellings were discovered. Particularly useful is
the balance of overall matter dependencies in $r_{s*}$ and $D_{a*}$
which helps account for the  $\Omega_M$-$\Omega_\Lambda$ degeneracy in
flat models.  At the same time, however, it is remarkable that the
admittedly simple arguments of this work yield such a decisive
cosmological indicator.
Within the next few years, NASA's MAP satellite data should give $\ell_{peak}$ 
to cosmic-variance levels. This measurement will burn a sharp line of
possible worlds across the ($\Omega_M,\Omega_K$) plane.

\acknowledgments
This letter was begun at Cambridge as part of DAMTP's 
tripos.  TM thanks Ofer Lahav and Daniel Wesley for early
guidance, David Hogg for cosmography computer code, and
Lyman Page and James Peebles for suggestions regarding the final draft. TM
is an NSF Graduate Research Fellow and is supported by NSF grant 
PHY-0099493.


\begin{thebibliography}{99}
  \bibitem[Bond et al. (1994)]{bond} Bond, J.R., Crittenden, R., Davis, R.L.,
Efstathiou, G., \& Steinhardt, P.J. 1994, Phys. Rev. Lett., 72, 13
  \bibitem[Burles, Nollett, \& Turner (2001)]{burles}Burles, S., Nollett, K.M.,
\& Turner, M.S. 2001, ApJ, 552, L1 
\bibitem[Cornish (2000)]{cornish}Cornish, N.J. 2000, preprint(astro-ph/0005261)
  \bibitem[Dodelson \& Knox\ (2000)]{dodelson}Dodelson, S., \& Knox, L.
2000, Phys. Rev. Lett., 84, 3523
  \bibitem[Doroshkevich, Zel'dovich, \& Syunyaev (1978)]{doroshkevich}
Doroshkevich, A.G., Zel'dovich, Ya.B., \& Syunyaev, R.A. 1978, Sov. Astron.,
22, 523
  \bibitem[Efstathiou \& Bond (1999)]{efstathiou}Efstathiou, G., \& Bond, J.R.
1999, MNRAS, 304, 75 
  \bibitem[Fixen et al. (1997)]{fixen} Fixen, D.J., Hinshaw, G., Bennet, C.L.,
and Mather, J.C. 1997, ApJ, 486, 623 
  \bibitem[Freedman et al. (2001)]{freedman_1} Freedman, W.L., et al. 
2001, ApJ, 553, 47 
  \bibitem[Halverson et al. (2001)]{halverson} Halverson, N.W., et al. 
2001, ApJ, submitted (astro-ph/0104489)
  \bibitem[Hogg (1999)]{hogg_1} Hogg, D.W. 1999, preprint(astro-ph/9905116)
  \bibitem[Hu \& Sugiyama\ (1996)]{hu_2} Hu, W., \& Sugiyama, N. 1996,
ApJ, 471, 542
  \bibitem[Hu et al. (2001)]{hu_3} Hu, W., Fukugita, M., Zaldarriaga, M., \&
Tegmark, M. 2001, ApJ, 549, 669 
  \bibitem[Jaffe et al. (2000)]{jaffe_1} Jaffe, A.H., et al. 2000,
Phys. Rev. Lett., 86, 3475 
  \bibitem[Kamionkowski, Spergel, \& Sugiyama (1994)]{kamionkowski_1} 
Kamionkowski, M., Spergel, D.N., \& Sugiyama, N. 1994, ApJ, 426, L57 
  \bibitem[Knox \& Page\ (2000)]{knox_1} Knox, L., \& Page, L.A. 2000,
Phys. Rev. Lett., 85, 1366 
  \bibitem[Knox, Christensen, \& Skordis  (2001)]{knox_2} Knox, L.,
Christensen, N., \& Skordis, C. 2001, ApJ, 563, L95
  \bibitem[Lee et al. (2001)]{lee} Lee, A.T., et al. 2001, ApJ, 561, L1 
  \bibitem[Miller et al. (1999)]{amber}Miller, A.D., et al. 1999, ApJ, 
524, L1 
  \bibitem[Netterfield et al. (2001)]{netterfield}Netterfield, C.B.,
et al. 2001, ApJ, in press (astro-ph/0104460)
  \bibitem[Page (2000)]{page_1}Page, L.A. 2000, in Proc. IAU Symposium 201,
ed. A. Lasenby \& A. Wilkinson, in press (astro-ph/0012214)
  \bibitem[Peebles (1993)]{peebles} Peebles, P.J.E. 1993, Principles of
Physical Cosmology, (Princeton University Press)
  \bibitem[Pryke et al. (2001)]{pryke}Pryke, C., Halverson, N.W., Leitch,
E.M., Kovac, J., Carlstrom, J.E., Holzapfel, W.L., \& Dragovan, M.  2001, 
ApJ, submitted (astro-ph/0104490)
  \bibitem[Seljak \& Zaldarriaga\ (1996)]{seljak_CMBFAST}Seljak, U., \&
Zaldarriaga, M. 1996, ApJ, 469, 437 
  \bibitem[Stompor et al. (2001)]{stompor}Stompor, R., et al. 
2001, ApJ, 561, L7 
  \bibitem[Weinberg (2000)]{weinberg_1}Weinberg, S. 2000, Phys. Rev. D,
62, 127302 
\end{thebibliography}
\end{document}